\documentclass[fleqn,usenatbib]{mnras}

\DeclareRobustCommand{\VAN}[3]{#2}
\let\VANthebibliography\thebibliography
\def\thebibliography{\DeclareRobustCommand{\VAN}[3]{##3}\VANthebibliography}

\usepackage[T1]{fontenc}
\usepackage{lmodern}
\usepackage{graphicx}	
\usepackage{amsmath}	
\usepackage{amssymb}	
\usepackage{amsfonts}
\usepackage{caption}
\usepackage{subcaption}
\usepackage{xcolor}

\usepackage{scalerel,tikz}
\usetikzlibrary{svg.path}
\definecolor{orcidlogocol}{HTML}{A6CE39}
\tikzset{orcidlogo/.pic={\fill[orcidlogocol] svg{M256,128c0,70.7-57.3,128-128,128C57.3,256,0,198.7,0,128C0,57.3,57.3,0,128,0C198.7,0,256,57.3,256,128z}; \fill[white] svg{M86.3,186.2H70.9V79.1h15.4v48.4V186.2z} svg{M108.9,79.1h41.6c39.6,0,57,28.3,57,53.6c0,27.5-21.5,53.6-56.8,53.6h-41.8V79.1z M124.3,172.4h24.5c34.9,0,42.9-26.5,42.9-39.7c0-21.5-13.7-39.7-43.7-39.7h-23.7V172.4z} svg{M88.7,56.8c0,5.5-4.5,10.1-10.1,10.1c-5.6,0-10.1-4.6-10.1-10.1c0-5.6,4.5-10.1,10.1-10.1C84.2,46.7,88.7,51.3,88.7,56.8z};}}
\newcommand\orcidicon[1]{\href{https://orcid.org/#1}{\mbox{\scalerel*{
\begin{tikzpicture}[yscale=-1,transform shape]\pic{orcidlogo};
\end{tikzpicture}}{|}}}}

\newcommand\hi{\mbox{\sc Hi}}
\newcommand\hii{\mbox{\sc Hii}}
\newcommand{\msun}{\mathrm{M}_{\odot}}
\newcommand{\Myr}{\mathrm{Myr}} 
\newcommand{\pc}{{\mathrm{pc}}}
\newcommand{\cs}{c_\mathrm{s}}
\newcommand{\va}{v_\mathrm{A}}
\newcommand{\sigsfr}{\Sigma_\mathrm{SFR}}
\newcommand{\sigsfrunit}{\msun\,\mathrm{yr^{-1}\,kpc^{-2}}}
\newcommand{\vff}{{f_\mathrm{WNM}}}
\newcommand{\mach}{\mathcal{M}}
\newcommand{\sigr}{\sigma_{\rho/\rho_0}}
\newcommand{\sigv}{\sigma_v}

\title[Turbulence in the ISM of TIGRESS]{Turbulence driving in a star-forming Milky-Way-type galaxy}

\author[Gerrard \& Federrath]{
Isabella~A.~Gerrard$^{\orcidicon{0000-0002-1995-6198}\,1}$ \& Christoph~Federrath$^{\orcidicon{0000-0002-0706-2306}\,1}$\thanks{Email:~\href{mailto:christoph.federrath@anu.edu.au}{christoph.federrath@anu.edu.au}}
\\
$^{1}$Research School of Astronomy and Astrophysics, Australian National University, ACT 2611, Australia
}

\pubyear{2025}

\begin{document}
\label{firstpage}
\pagerange{\pageref{firstpage}--\pageref{lastpage}}
\maketitle

\begin{abstract}
The life-cycle, structure, and dynamics of the interstellar medium (ISM) is regulated by turbulence. Complex physical processes, including supernova (SN) explosions, shear, and gravitational collapse, drive and maintain turbulence, but it is still an open question what turbulence driving mode is primarily excited by these different mechanisms. The turbulence driving parameter, $b$, can be used to quantify the ratio of solenoidal to compressive modes in the acceleration field that drives the turbulence. Compressive driving is characterised by $b\sim1$, while purely solenoidal driving gives $b\sim0.3$. To quantify the turbulence in the galactic ISM, we investigate the time evolution of $b$, as well as the turbulent Mach number, and plasma $\beta$ (thermal-to-magnetic pressure ratio), and its correlation with star formation in the magnetised warm neutral medium (WNM) of the TIGRESS shearing-box simulations of a $\sim$~kpc-sized patch of a Milky-Way-like galaxy, over a $100\,\Myr$ time period ($\sim$~half an orbital time). In this simulation the turbulence is driven by a combination of shear, gravitational collapse, and star formation feedback in the form of radiation and SNe. We find that the turbulence driving parameter fluctuates in time between $b\sim0.4$ and $b\sim1$. We find a time-dependent correlation of $b$ with star formation activity, such that high star formation rates follow about one turbulent turnover time ($\sim10\,\Myr$) after phases of highly compressive driving ($b>0.5$). About $20\,\Myr$ after the peak in star formation, type-B SN feedback drives up the WNM fraction and turbulent Mach numbers, and reduces plasma $\beta$ and the driving to $b\sim0.4-0.5$.
\end{abstract}



\section{Introduction}

Turbulence plays a crucial role in determining how the ISM evolves within galaxies. To understand the processes that control the ISM's life cycle, structure, and dynamics, it is essential to analyse the turbulence in a diverse range of galactic environments by comparing observations with theoretical models and simulations \citep[for a review, see][]{Burkhart2021}. Realistic, resolved computational models of the star-forming ISM, which include radiation feedback, SNe, star formation, magnetic fields, chemistry, and large-scale galactic motions, continue to progress our understanding of the intricate interplay between these processes and their role in shaping the ISM that we observe. Statistics that capture the behaviour of the turbulence produced by these complex physical mechanisms are useful probes when comparing across environments, simulations, and observations. 

One such quantity is the \emph{turbulence driving parameter} \citep{FederrathKlessenSchmidt2008,FederrathDuvalKlessenSchmidtMacLow2010}, which describes the ratio of compressive to solenoidal modes in the acceleration field that drives turbulence. This parameter encapsulates information about how energy is being injected into the ISM via physical processes such as star formation feedback or galaxy dynamics. The studies by \citet{PadoanNordlund2011}, \citet{PriceFederrathBrunt2011}, \citet{KonstandinEtAl2012ApJ}, \citet{MolinaEtAl2012}, \citet{NolanFederrathSutherland2015}, \citet{FederrathBanerjee2015}, and \citet{KainulainenFederrath2017} have shown that the width of the density PDF is proportional to the turbulent (sonic) Mach number, and that for an isothermal, supersonic, magnetised gas,
\begin{equation}
\sigr = \frac{b \mach}{\sqrt{1+1/\beta}} \;\;\implies\;\; b = \sigr \mach^{-1} (1+1/\beta)^{1/2},
\label{eq:b}
\end{equation}
where $\sigr$ is the standard deviation of the 3D turbulent density field ($\rho$), scaled by the mean density ($\rho_0$), $\mach=\sigv/\cs$ is the 3D turbulent sonic Mach number (ratio of turbulent velocity dispersion to sound speed), plasma $\beta$ is the ratio of thermal to magnetic pressure, and $b$ is the turbulence driving parameter \citep{FederrathKlessenSchmidt2008,FederrathDuvalKlessenSchmidtMacLow2010,PadoanNordlund2011,MolinaEtAl2012}. This relationship has been extensively studied in simulations of driven turbulence \citep[e.g.,][]{FederrathKlessenSchmidt2008,FederrathDuvalKlessenSchmidtMacLow2010,PriceFederrathBrunt2011,MolinaEtAl2012,NolanFederrathSutherland2015,FederrathBanerjee2015,BeattieEtAl2021}. Methods for measuring $\sigr$ and $\mach$, and therefore allowing a reconstruction of $b$ in an observational context have been developed in \citet{BruntFederrathPrice2010a,BruntFederrathPrice2010b}, \citet{GinsburgFederrathDarling2013}, \citet{KainulainenFederrathHenning2014}, \citet{BruntFederrath2014}, \citet{StewartFederrath2022}, \citet{GerrardEtAl2023,GerrardEtAl2024} and \citet{NarayanTritsisFederrath2025}. Further recent studies related to Eq.~(\ref{eq:b}) include 
\citet{BeattieEtAl2021}, \citet{DhawalikarEtAl2022}, and \citet{BandyopadhyayBeattieBhattacharjee2025}.

What is yet to be explored is the evolution of the turbulence driving parameter over time. Moreover, the majority of the previously mentioned theoretical studies have examined the behaviour of a simulated gas when the mixture of driving modes is explicitly set as an input parameter in the simulation \citep[cf.~fig.~8 in][]{FederrathDuvalKlessenSchmidtMacLow2010}. Here we take advantage of the unique opportunity that the TIGRESS simulation suite provides to explore the evolution of $b$ in a kpc-sized patch of a Milky-Way-like ISM, where multiple complex physical processes self-consistently drive and maintain turbulence, and that includes complex phase structure as well as magnetic fields. We study how the turbulence in the diffuse ISM may correlate (or not) with star formation and the associated SN feedback.

Section~\ref{sec:methods} summarised the TIGRESS simulation used here and the methods to isolate and measure the relevant turbulent quantities that determine the turbulence driving parameter in Eq.~(\ref{eq:b}). In Sec.~\ref{sec:results} we show the spatial distribution of the diffuse ISM in the simulations and present the main turbulence analyses, as well as a quantification of the time evolution of the turbulence driving parameter and its correlation with star formation activity. Section~\ref{sec:con} provides a summary and conclusions.


\section{Methods} \label{sec:methods}

In this section we describe the TIGRESS simulations and our methods to analyse the turbulence of the $\hi$ gas therein. Our aim is to investigate the behaviour of the density dispersion, turbulent sonic Mach number, and ultimately the turbulence driving parameter over time, and the underlying mechanism that dominates the driving of turbulence. As opposed to idealised simulations using artificial Fourier driving, there is no explicit prescription of the mixture of driving modes in TIGRESS, and therefore the driving parameter is a natural outcome of the physical processes \citep{Elmegreen2009,FederrathEtAl2017iaus} implemented in the simulation, i.e., a combination of galaxy rotation, shear, gravity, accretion, and radiation + SN feedback. 


\subsection{TIGRESS simulations} \label{sec:sims}

We use the public data release\footnote{\url{https://princetonuniversity.github.io/astro-tigress/intro.html}} of the TIGRESS simulation suite \citep{KimOstriker2017}. The details of the methods used in these simulations are presented in \citet{KimOstriker2017}, and further modifications to the star formation and accretion techniques can be found in \citet{KimEtAl2020}. Here we give a brief summary of the main ingredients. 

The TIGRESS framework solves the ideal MHD equations using \texttt{Athena} \citep{StoneEtAl2008,StoneGardiner2009} in a shearing-box \citep{StoneGardiner2010} with a galactic rotation speed of $\Omega_0=28\,\mathrm{km\,s^{-1}\,kpc^{-1}}$, assuming a flat rotation curve. Self-gravity is solved using a fast Fourier transform (FFT) method under horizontal shearing-periodic and vertically open boundary conditions \citep{KoyamaOstriker2009}. The vertical gravity of the stellar disc and dark matter halo is included as a fixed external potential \citep{ZhangEtAl2013}. A tabulated cooling function following \citet{KoyamaInutsuka2002} at $T<10^{4.2}\,\mathrm{K}$ and \citet{SutherlandDopita1993} at $T>10^{4.2}\,\mathrm{K}$ is used to solve optically thin cooling for the full range of gas temperatures. When gas cools and collapses into dense, gravitationally bound structures (defined by a density threshold and converging flow conditions), sink particles are created to follow further accretion and feedback \citep{GongOstriker2013,KimEtAl2020}. Each sink particle represents a star cluster whose FUV luminosity and SN rate are calculated based on the STARBURST99 simple stellar population synthesis model \citep{LeithererEtAl1999}. The photoelectric heating rate scales linearly with the interstellar radiation field calculated with a globally attenuated total FUV luminosity from star clusters \citep{OstrikerKim2022}. SNe are probabilistically assigned for a given star cluster's SN rate; 1/3 of them are exploded as runaways, and the rest are exploded at the sink particles' location.

Here we use the highest-resolution solar neighbourhood model \citep[R8-2pc in their naming convention or R8-Z1 in][]{GongEtAl2020}. The simulation covers a volume $V=L_x\times L_y\times L_z=1024\times1024\times7168\,\pc^3$ of uniform resolution with a cell size of $\Delta x = 2\,\pc$. We also compare this with a lower-resolution version ($\Delta x = 4\,\pc$), i.e., a version of the TIGRESS simulation at double the cell size. This is shown in Appendix~\ref{app:resolution}. We find that the resulting turbulence driving parameter is relatively insensitive to changes in the numerical resolution.

In order to study the time evolution, we use 11~snapshots over $\sim 100\,\Myr$ (in original simulation time: $t\sim274-372\,\Myr$), equally spaced by $\sim 9.8\,\Myr$. The total simulation time corresponds to approximately half an orbital time of the galaxy. The simulation self-consistently forms star clusters in a turbulent, multiphase, magnetised ISM. The turbulent, thermal, and magnetic pressures of the ISM reach a quasi-steady saturation state \citep{KimChoiFlauger2019}, controlled by the interplay of SN feedback, FUV radiation, and magnetic fields \citep{OstrikerKim2022}. The emerging star formation rate in the simulation agrees well with observational estimates, with values of $\sigsfr\sim10^{-3}-10^{-2}\,\sigsfrunit$ \citep[e.g.,][]{FuchsEtAl2009,ZariFrankelRix2023}.


\subsection{Identifying the warm neutral medium (WNM)} \label{sec:wnm}

In this study we consider only the WNM, because it is the most volume-filling of the $\hi$ phases (near the mid-plane), and is approximately isothermal. We can also directly observe the WNM via 21\,cm emission, and compare our results in these simulations with previous measurements of turbulence statistics observed in the WNM \citep[e.g.,][]{MarchalMivilleDeschenes2021,GerrardEtAl2023,GerrardEtAl2024}. To separate out the WNM, we mask the raw data with a temperature range of $5000 \leq T/\mathrm{K} \leq 8000$ in accordance with the definition of the WNM in \citet{McClureGriffithsEtAl2023}. With a mean particle weight of $1.3\,m_\mathrm{H}$, where $m_\mathrm{H}$ is the mass of a hydrogen atom, the sound speed of this gas is approximately \mbox{$\cs=8\pm1\,\mathrm{km\,s^{-1}}$} \citep[e.g., eq.~8 in][]{FederrathOffner2025}.


\subsection{Separating turbulence from non-turbulent contributions} \label{sec:turbiso}

Eq.~(\ref{eq:b}) strictly applies only to turbulent gas. Thus, we need to separate the turbulent density and velocity fluctuations from the systematic (non-turbulent) contributions, such as the vertical density stratification of the disc, and the shearing motions in the plane of the galaxy. In order to isolate the turbulent contributions to the fluctuations in the density, velocity, and magnetic field, we smooth these fields. The smoothed fields represent the non-turbulent contributions. The smoothing is done with a low-pass filtering operation\footnote{Filtering necessarily modifies the mode distribution of the fields, but this is intentional: applying Eq.~(\ref{eq:b}) requires isolating purely turbulent fluctuations. While the filtering can change the divergence and/or curl compared to the original fields, this is a feature rather than a drawback, as some modes are non-turbulent in origin (e.g., density gradients from gravitational collapse or vertical stratification) and must be removed.} that uses a Gaussian kernel with an $L_\mathrm{kernel}=100\,\pc$ FWHM. We chose $100\,\pc$ as it is often considered a reasonably characteristic injection scale of the turbulence in disk galaxies and is of the order of the molecular disk scale height\footnote{We note that the turbulent turnover time of gas with a typical WNM velocity dispersion of $\sigv\sim10-20\,\mathrm{km\,s^{-1}}$ is \mbox{$t_\mathrm{turb}=L_\mathrm{kernel}/\sigv\sim5-10\,\Myr$} on $L_\mathrm{kernel}=100\,\pc$ scales.}. However, we have investigated the dependence of our results on this choice, by comparing different kernel sizes in Appendix~\ref{app:kernelsize}, and find that $b$ is largely insensitive to the particular choice of kernel size. Variations on scales larger than $L_\mathrm{kernel}$ are considered non-turbulent fluctuations with respect to the size of the full region (here $1024\,\pc$ in the disc plane direction and up to $\sim\pm2\,\mathrm{kpc}$ in the vertical direction, beyond which the WNM fraction falls drastically as we will see below), while scales smaller than the filter FWHM are considered turbulent \citep{FederrathEtAl2016,StewartFederrath2022,GerrardEtAl2024}.

In general, the turbulence isolation proceeds as follows. Assuming a quantity $\mathcal{Q}$ (such as a velocity component of the gas), the filtering provides a smooth 3D version of $\mathcal{Q}$, which we denote $\mathcal{Q}_\mathrm{smooth}$. Based on this, the turbulent fluctuations ($\mathcal{Q}_\mathrm{turb}$) in the original field $\mathcal{Q}$ are obtained by subtracting the smooth field, as
\begin{equation} \label{eq:smoothing}
\mathcal{Q}_\mathrm{turb} = \mathcal{Q} - \mathcal{Q}_\mathrm{smooth}.
\end{equation}

In the following subsections, we specify which fields undergo this turbulence-isolation procedure. As a technical note, we mention that the original field $\mathcal{Q}$ may be sparse in that it contains NaN in places without WNM. In order to allow for Gaussian smoothing of such a sparse field, we use a Gaussian filter that ignores the NaN cells, by correctly weighting the contributions of non-NaN cells within the filter kernel, which is achieved by using the \texttt{scipy generic\_filter} function.


\subsubsection{Turbulent density dispersion ($\sigr$)} \label{sec:sigr}

To obtain the turbulent density field (with non-turbulent contributions removed; e.g., gradients introduced by the vertical structure of the disc) we use $\mathcal{Q} = \log_{10}(\rho/\rho_0)$ in Eq.~(\ref{eq:smoothing}), where $\rho_0$ is the mean density of the WMN (i.e., NaNs excluded). The logarithmic version of $\rho$ is used in the smoothing, because $\rho$ itself varies over many orders of magnitude, while $\log_{10}(\rho/\rho_0)$ does not, and therefore provides a more robust quantity with respect to extreme values. Moreover, $\log_{10}(\rho/\rho_0)$ follows a Gaussian distribution, making this quantity the more relevant quantity for turbulence isolation \citep[e.g.,][]{Vazquez1994,KritsukEtAl2007,FederrathDuvalKlessenSchmidtMacLow2010}.

After applying Eq.~(\ref{eq:smoothing}) we have the turbulent logarithmic density fluctuations, $(\log_{10}(\rho/\rho_0))_\mathrm{turb}$, which are transformed back to the normal (linear) turbulent density fluctuations via
\begin{equation}
(\rho/\rho_0)_\mathrm{turb} = 10^{\mathcal{Q}_\mathrm{turb}} = 10^{(\log_{10}(\rho/\rho_0))_\mathrm{turb}}.
\end{equation}

Finally, we compute the standard deviation of $(\rho/\rho_0)_\mathrm{turb}$ (again excluding NaN cells) to get the dispersion of the normalised turbulent density field, denoted $\sigr$ in Eq.~(\ref{eq:b}).


\subsubsection{Turbulent sonic Mach number (\texorpdfstring{$\mach$}{Mach})} \label{sec:mach}

To construct the turbulent sonic Mach number in Eq.~(\ref{eq:b}), we need the turbulent velocity field and the sound speed of the gas (the latter we have; cf.~Sec.~\ref{sec:wnm}). In order to obtain the turbulent velocity field, we apply the turbulence-isolation method (Sec.~\ref{sec:turbiso}) to each component of the velocity field, i.e., $\mathcal{Q} = v_x, v_y, v_z$ in Eq.~(\ref{eq:smoothing}). Using the turbulent velocity components, we can now define the local (cell-based) Mach number in each spatial direction,
\begin{align} \label{eq:mach_xyz}
    M_x,\,M_y,\,M_z =&\,\frac{v_x}{\cs},\,\frac{v_y}{\cs},\,\frac{v_z}{\cs}.
\end{align}
Finally, the turbulent sonic Mach number in Eq.~(\ref{eq:b}) is computed from the standard deviation of each Mach number component, and combining them to obtain the 3D turbulent Mach number,
\begin{equation} \label{eq:mach}
    \mach = \left(\sigma_{M_x}^2 + \sigma_{M_y}^2 + \sigma_{M_z}^2\right)^{1/2}.
\end{equation}


\begin{figure*}
\centering
\includegraphics[width=0.99\linewidth]{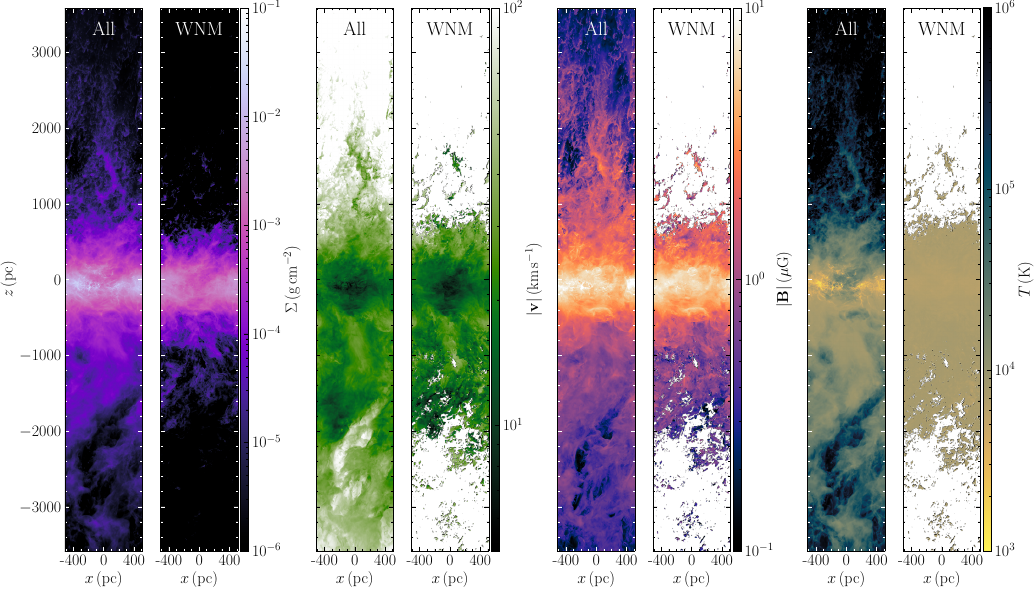}
\caption{An example snapshot (at $t=293\,\Myr$) of the TIGRESS simulation domain. The panels show 4~pairs (from left to right) corresponding to the column density, and the mass-weighted projections of the velocity, magnetic field, and temperature, along the $y$-axis (the galactic plane is the $z=0$ plane). The left and right component of each pair shows all gas and WNM gas (as defined in Sec.~\ref{sec:wnm}), respectively (as labelled). The first 3~pairs (column density, velocity, and magnetic field) are the quantities from which we construct the turbulent density dispersion, sonic Mach number, and plasma $\beta$, and finally the turbulence driving parameter $b$ by combining them.}
\label{fig:simbox}
\end{figure*}

\subsubsection{Plasma beta (\texorpdfstring{$\beta$}{beta})} \label{sec:plasma_beta}

Plasma beta in Eq.~(\ref{eq:b}) is defined as the ratio of the thermal to magnetic pressure, which can be expressed as a ratio of speeds \citep{FederrathKlessen2012},
\begin{equation} \label{eq:beta}
\beta = 2\,\frac{\cs^2}{\va^2},  
\end{equation}
where $\va=B/(4\pi\rho)^{1/2}$ is the Alfv\'en speed and $B=\vert\mathbf{B}\vert$ is the magnetic field strength. Eq.~(\ref{eq:beta}) holds for approximately isothermal gas, which is reasonably fulfilled given the WNM phase mask defined in Sec.~\ref{sec:wnm}.

Since the turbulent component of the magnetic field is primarily relevant for Eq.~(\ref{eq:b}), we apply the turbulence-isolation method (Sec.~\ref{sec:turbiso}) to the individual components of $\mathbf{B}$, such that $\mathcal{Q} = B_x, B_y, B_z$ in Eq.~(\ref{eq:smoothing}), and then use $\vert\mathbf{B}_\mathrm{turb}\vert$ to construct the Alfv\'en speed $\va$, to finally obtain $\beta = \beta_\mathrm{turb}$ via Eq.~(\ref{eq:beta}).

We note that the plasma-$\beta$ correction term $(1+1/\beta)^{1/2}$ in Eq.~(\ref{eq:b}), derived in \citet{MolinaEtAl2012}, assumes that the magnetic field scales with the density as $B\propto\rho^{1/2}$. We show that the simulation follows this relation to reasonable approximation in Appendix~\ref{app:B_rho}.

We further note that, provided the mean magnetic field is relatively weak (Alfv\'en Mach number $\gtrsim 2$ with respect to the mean field), turbulent tangling yields field fluctuations comparable to the mean field \citep{Federrath2016jpp,BeattieFederrathSeta2020}. In this regime, there is no significant distinction between mean and turbulent magnetic fields, which applies for the simulation case studied here (cf.~Fig.~\ref{fig:B_rho}). Extensions to the strong-field case have been developed \citep{BeattieEtAl2021}, but are not required here; we therefore adopt the \citet{MolinaEtAl2012} scaling of $B\propto\rho^{1/2}$ for the magnetic-field correction term in Eq.~(\ref{eq:b}) (see Appendix~\ref{app:B_rho}).

\section{Results} \label{sec:results}

In this section we describe the results of the turbulence analysis, with a particular focus on the turbulence driving mode ($b$ in Eq.~\ref{eq:b}), over the course of the time evolution of the simulation, and explore whether, and if so, how these turbulence quantities correlate with the formation of stars. 

\subsection{Spatial distribution of turbulent gas} \label{sec:spatial}

Figure~\ref{fig:simbox} shows the column density, velocity, magnetic field, and temperature in a snapshot of the simulation at $t=293\,\Myr$, presenting a side-on view along the galactic plane (with the plane located at $z=0$). All quantities show typical values associated with a Milky-Way-type galaxy, with a magnetic field of $\sim0.1-10\,\mu\mathrm{G}$. We note that the WNM gas covers the disc and extends vertically up to $\sim\pm2\,\mathrm{kpc}$. The WNM gas covers a relatively narrow temperature range as per definition in Sec.~\ref{sec:wnm} and therefore can be approximated as nearly isothermal, which facilitates the applicability of Eq.~(\ref{eq:b}). Using the WNM dataset as a basis, we then follow the methods described in Sec.~\ref{sec:turbiso}, in order to isolate the turbulent fluctuations in the WNM gas. The results of this turbulence isolation process are discussed in the next section.

\subsection{Turbulence isolation}

\begin{figure}
\centering
\includegraphics[width=0.95\linewidth]{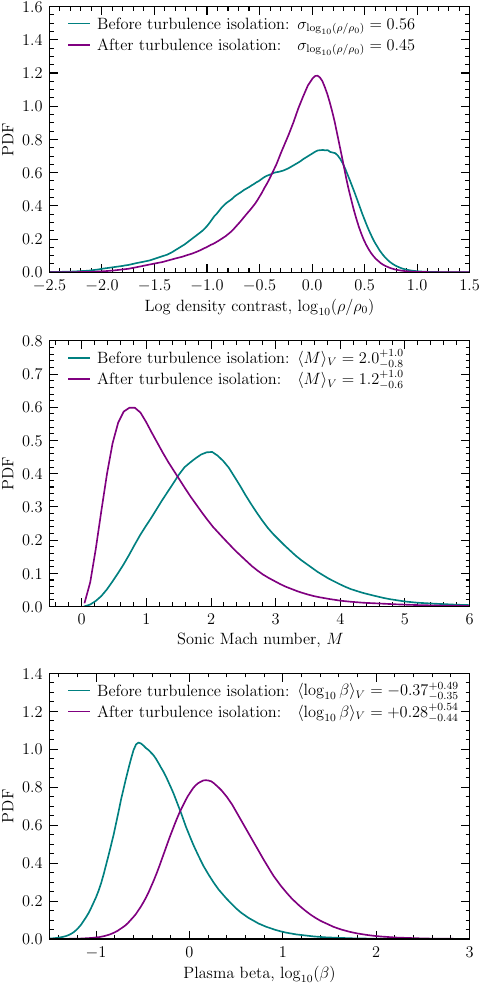}
\caption{PDFs of the three turbulence analysis quantities that constitute $b$ (Eq.~\ref{eq:b}), computed from the WNM data shown in Fig.~\ref{fig:simbox}. The panels show the PDFs of the log-density field $\log_{10}(\rho/\rho_0)$ (top), the sonic Mach number $M=(M_x^2+M_y^2+M_z^2)^{1/2}$ (middle), and the plasma $\beta$ (bottom), before (teal) and after (purple) turbulence isolation (see Sec.~\ref{sec:turbiso}). Turbulence isolation yields a reduction in the standard deviation of the density and in the average of the sonic Mach number, and an increase in the average $\beta$, as expected when the respective non-turbulent contributions to the fields is removed.}
\label{fig:pdfs}
\end{figure}

The first step in applying Eq.~(\ref{eq:b}) is to isolate the turbulent fluctuations and separate them from non-turbulent contributions. The methods for doing this are outlined in Sec.~\ref{sec:turbiso}. Figure~\ref{fig:pdfs} shows the results of the turbulence isolation, which is applied to all components that enter Eq.~(\ref{eq:b}), i.e., the density, sonic Mach number, and plasma $\beta$. A corresponding visualisation of the three turbulent fields is presented in Appendix~\ref{app:turb_fields}.

As discussed in Sec.~\ref{sec:sigr}, the turbulence-isolation method is applied to the normalised density field and its effects are shown in Fig.~\ref{fig:pdfs} (top panel), with the PDF of $\log_{10}(\rho/\rho_0)$ before and after turbulence isolation. We see that the shape of the PDF becomes more log-normal and the standard deviation (denoted $\sigma$ on the figure) of the density field is reduced after turbulence isolation, i.e., the remaining fluctuations in $\rho$ are primarily of turbulent origin (with the non-turbulent contributions, such as the vertical disc stratification, removed).

The middle panel of Fig.~\ref{fig:pdfs} shows the same comparison (before and after turbulence isolation) for the absolute value of the local sonic Mach number in each cell, i.e., $M=(M_x^2+M_y^2+M_z^2)^{1/2}$ (cf.~Sec.~\ref{sec:mach}). As expected, the turbulence isolation yields a reduction in the average $M$, i.e., non-turbulent contributions (such as large-scale shear) were successfully filtered out of the velocity field. Typical turbulent Mach numbers of the WNM are $\mach\sim1-3$.

Similarly, the bottom panel of Fig.~\ref{fig:pdfs} shows the same for plasma $\beta$, where the underlying magnetic field $\mathbf{B}$ has been filtered (cf.~Sec.~\ref{sec:plasma_beta}). We find that the average value of $\beta$ increases, and typical values are of the order of $\beta\sim1-10$ after turbulence isolation in the WNM. We note that applying the turbulence isolation on the magnetic field is necessary in order to obtain the isotropic plasma-$\beta$ contribution to Eq.~(\ref{eq:b}). While the mean field is also associated with magnetic pressure, it corresponds to the large-scale, non-isotropic component of $\beta$, which does not directly enter the relation for $b$ established in Eq.~(\ref{eq:b}) -- it is the unordered, turbulent component that matters in this context.

Now that we have isolated the turbulent contributions, we can compute the respective statistical quantities entering Eq.~(\ref{eq:b}), i.e., $\sigr$, $\mach$, and $\beta$, from their turbulence-isolated fields.

\subsection{Turbulence and its instantaneous correlation with star formation} \label{sec:sfr}

\begin{figure}
\centering
\includegraphics[width=0.95\linewidth]{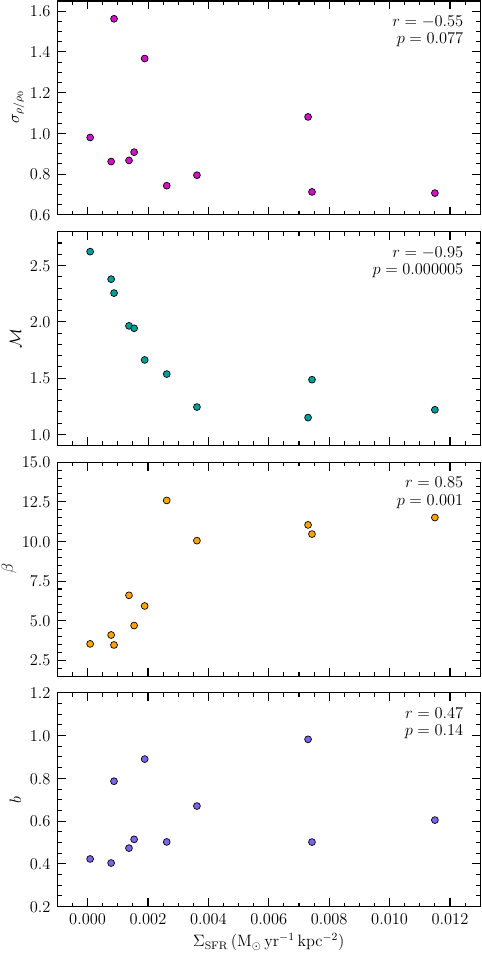}
\caption{The four turbulence analysis quantities as a function of star formation surface density ($\sigsfr$). The panels from top to bottom show: the density dispersion ($\sigr$), the sonic Mach number ($\mach$), the ratio of thermal to magnetic pressure ($\beta$), and the turbulence driving parameter ($b$). The Spearman rank correlation coefficient $r$ and the corresponding $p$-value are listed in the top-right corner of each panel. Only $\mach$ and $\beta$ show a statistically significant negative and positive correlation, respectively, with $\sigsfr$, while a relatively weak positive correlation is seen between $\sigr$ and $\sigsfr$, and $b$ and $\sigsfr$, which may or may not be regarded significant ($p\sim0.1$).}
\label{fig:turbsfrcorr}
\end{figure}

Here we summarise the main results of the turbulence analysis. Figure~\ref{fig:turbsfrcorr} shows $\sigr$, $\mach$, $\beta$, and $b$ as a function of the star formation rate surface density ($\sigsfr$), i.e., each point represents a particular snapshot (in the time range of $274-372\,\Myr$) of the simulation. We first focus on the distribution of the turbulence quantities before studying any correlations with $\sigsfr$.

\subsubsection{The range of turbulence conditions in the WNM}

We find that $\sigr$, $\mach$, $\beta$, and $b$ occupy a reasonable parameter space for the present simulations and associated physical conditions. The Mach number is in the range $\mach\sim1-3$, i.e., transsonic to slightly supersonic over the course of the simulation, on the scale it is measured ($\sim100\,\pc$; cf.~Sec.~\ref{sec:turbiso})\footnote{We note that the velocity dispersion scales with the length scale as a power law with an exponent of $p\sim0.5$ in the supersonic regime, and $p\sim 0.4$ in the subsonic regime \citep[e.g.][]{Larson1981,OssenkopfMacLow2002,HeyerBrunt2004,RomanDuvalEtAl2010,FederrathEtAl2021}. Thus, the Mach numbers we find here apply to WNM gas on $\sim100\,\pc$ scales; the corresponding Mach numbers on smaller and larger scales are respectively smaller and larger than that -- see Appendix~\ref{app:kernelsize}.}

Plasma beta is in the range $\beta\sim3-13$, i.e., the thermal pressure is larger than the magnetic pressure by factors of a few up to an order of magnitude, on average, which is in a reasonable range considering the high temperatures of the WNM ($T\sim5000-8000\,\mathrm{K}$). Plasma beta is expected to drop below unity in the cold, molecular, star-forming phase of the ISM \citep{FalgaroneEtAl2008}.

Finally, we find that the turbulence driving parameter is in the range $b\sim 0.4-1.0$, i.e., covering almost the entire possible range, from purely solenoidal ($b\sim1/3$) to purely compressive ($b\sim1$). All snapshots exhibit $b\ge0.4$, which indicates a dominance of compressive driving modes \citep[$b=0.38$ is the natural driving mixture, and anything above that has more compressive than solenoidal modes in the driving field; see fig.~8 in][]{FederrathDuvalKlessenSchmidtMacLow2010}.

\subsubsection{Instantaneous correlation of turbulence with star formation}

We now study the instantaneous correlations\footnote{By `instantaneous correlation' we mean that two quantities being compared are evaluated at the same simulation time, i.e., each data point represents a simultaneous pair, e.g.~$(\sigsfr,\,b)$, taken at a given simulation snapshot. We emphasise this usage because some quantities may exhibit a time lag between cause (e.g., changes in SFR) and response (e.g., variations in $b$), or vice versa, which is not accounted for when considering an instantaneous correlation in time.} of $\sigr$, $\mach$, $\beta$, and $b$ with $\sigsfr$, shown by Spearman's rank correlation coefficients ($r$) and their associated $p$ values in Fig.~\ref{fig:turbsfrcorr}. We find a potentially weak, yet only marginally statistically significant correlation (cf.~$p\sim0.1$) between $\sigsfr$ and either the density dispersion $\sigr$ (top panel) or $b$ (bottom panel). In particular, if the weak positive correlation of $b$ with $\sigsfr$ is regarded statistically significant (depending on the choice of cutoff $p$), then this would hint towards more compressive driving being associated with higher star formation rates, but the trend and statistical significance are rather weak.

The Mach number and plasma beta show statistically significant correlations with $\sigsfr$, in that $\mach$ is anti-correlated and $\beta$ is positively correlated with $\sigsfr$. Thus, the turbulent Mach number tends to be lower when the star formation rate (SFR) is high, and vice versa\footnote{We note that $\mach$ is almost exclusively determined by the turbulent velocity dispersion, rather than the sound speed. The latter is nearly constant as a result of the nearly isothermal approximation of the WNM phase selection -- see Appendix~\ref{app:turbsfrcorr_cs_sigv}.}. This may seem counterintuitive at first, as high SFR is often associated with a high level of turbulence, characterised by high Mach numbers. However, as we will see below, it is the SN feedback that drives $\mach$ up, and that feedback is delayed with respect to star formation. The origin of the positive relationship between $\beta$ and $\sigsfr$ is not immediately clear, as star formation has an influence on both the thermal and magnetic pressure components that make up $\beta$. The positive relation implies that higher SFR would thus be associated with either higher thermal pressure and/or lower magnetic pressure. We will revisit this in the following discussion, where we investigate the time evolution of these quantities.

In summary, while Fig.~\ref{fig:turbsfrcorr} shows the instantaneous correction between star formation and turbulence, it does not take into account the time dependence and potential delays in each of these correlations. In particular, we expect there to be certain delays in the effects of a star formation event and its associated feedback from SNe, as most SNe (B-type stars) take $\sim10-30\,\Myr$ to evolve before they explode. Thus, it is useful to investigate the time evolution of the simulations and the response of the turbulence to star formation in a time-dependent fashion.

\subsection{Time evolution of turbulence and star formation} \label{sec:tevol}

\begin{figure}
\centering
\includegraphics[width=0.95\linewidth]{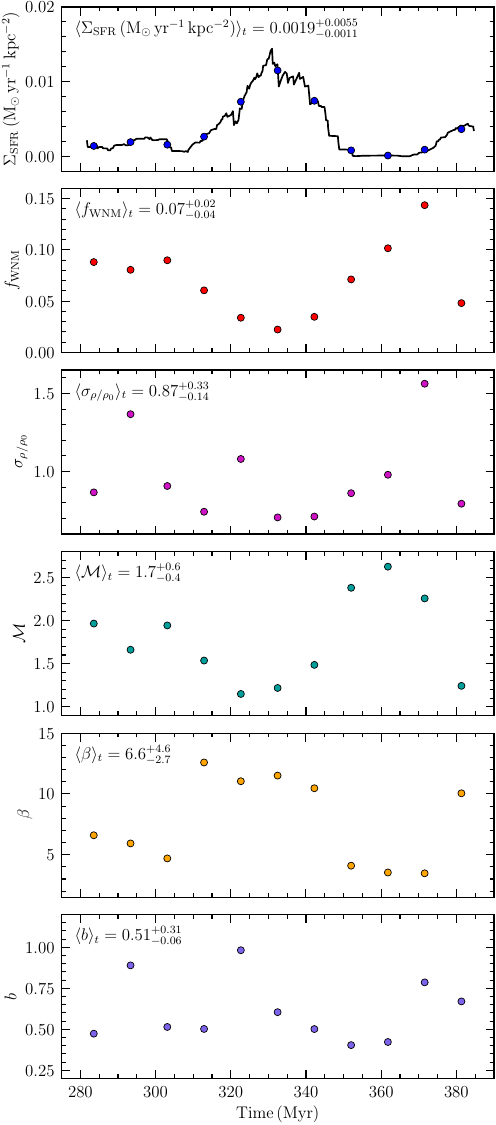} 
\caption{From top to bottom: time evolution of the star formation rate surface density, volume filling fraction of the WNM relative to the entire simulation volume, turbulent density dispersion, Mach number, plasma $\beta$, and turbulence driving parameter ($b$). Overall, we find evidence for a time-dependent correlation of $b$ with star formation, in that the phase leading up to the star formation peak is associated with increasing $b$ (i.e., more compressive driving), while the phase after star formation feedback shows a rise in $\vff$ and $\mach$, and an associated drop in $\beta$ and $b$ (down to a natural mixture of driving modes, $b\sim0.4$).}
\label{fig:tevol}
\end{figure}

The time evolution of the turbulence driving mode has not been explored in much detail before, except for one study, which has investigated the time evolution of $b$ during the expansion of an $\hii$ region \citep{MenonFederrathKuiper2020}. Therefore, we now explore the time evolution in the present simulations, with a particular focus on quantifying the amount of variation in the turbulence characteristics, as well as any potential correlation with star formation activity. The TIGRESS simulation suite provides 11~time instances separated by $\sim9.8\,\Myr$, i.e., a total of $\sim100\,\Myr$, representing roughly half an orbital period of the galaxy. Because the turbulence driving in these simulations is governed only by the physical processes implemented (e.g., star formation, SNe, radiation transfer, and shear), which self-consistently occur during the evolution of the simulation, there is no prescription for what the turbulence driving mode mixture should be at any given time.

Fig.~\ref{fig:tevol} shows the time evolution of (from top to bottom) the SFR surface density ($\sigsfr$), the volume filling fraction ($\vff$) of the WNM (i.e., the ratio of WNM to total simulation volume), and the 4~main turbulence analysis quantities ($\sigr$, $\mach$, $\beta$, and $b$). 

The top panel of Fig.~\ref{fig:tevol} shows $\sigsfr$, which has been tracked in the simulation with a finer time cadence (solid black line) than the full 3D snapshots (blue markers). A major star formation event occurs between $310 - 350\,\Myr$, peaking $t\sim330\,\Myr$ with $\sigsfr\sim 0.015\,\sigsfrunit$.

The $\vff$ of the WNM (second panel of Fig.~\ref{fig:tevol}) is approximately inversely proportional to $\sigsfr$, such that the $\vff$ is at its lowest when $\sigsfr$ is at its highest, followed by an increase in $\vff$ after star formation has ceased. This signifies that large fractions of WNM can reform out of the ionised gas about $30-40\,\Myr$ after the major star formation event, during which ionising radiation can transform major fractions of $\hi$ into $\hii$.

We also see the density dispersion ($\sigr$; third panel in Fig.~\ref{fig:tevol}) increasing during the built-up of the WNM after the star formation event, likely due to delayed SN feedback during that phase, driving local compressions in SN shells. Moreover, reduced star formation activity allows gas to regather gravitationally. It should be noted that the lowest density dispersion occurs at the peak of the star formation event when $\vff$ is also at a minimum, and similarly peaks when $\vff$ peaks. Apart from this, $\sigr$ can change significantly on shorter timescales ($\sim10\,\Myr$) than either $\sigsfr$ or $\vff$, likely due to individual SN explosions acting on those shorter timescales, creating variations in $\sigr$ of the order of $\pm50\%$.

In contrast, the Mach number ($\mach$; fourth panel in Fig.~\ref{fig:tevol}) evolves more smoothly in time, somewhat mimicking the functional form of $\vff$. During the star formation event, when $\vff$ is low, $\mach$ is also at a minimum, followed by a disturbance caused by delayed SN feedback, leading to an increase in $\mach$, and associated local compression of gas. We note that the changes in $\mach$ are not caused by changes in the temperature, but closely reflect changes in the turbulent velocity dispersion (cf.~Appendix~\ref{app:turbsfrcorr_cs_sigv}).

The ratio of thermal to magnetic pressure, plasma $\beta$ (fifth panel in Fig.~\ref{fig:tevol}) shows some signatures of the main star formation event, in that $\beta$ tends to be elevated during the phase of high star-formation activity, followed by a phase of lower $\beta$, which may be the result of local compression via delayed SN feedback. Considering a time delay of $\sim10-30\,\Myr$ (the time for type-B SNe to explode), we can associate high SN feedback activity with low $\beta$ and increased $\mach$, reflecting the driving of turbulence and the associated compression and/or tangling of the magnetic field in SN shocks. Indeed, $\mach$ is high and $\beta$ is low $\sim20\,\Myr$ after the star formation peak.

Finally, the bottom panel of Fig.~\ref{fig:tevol} shows the turbulence driving parameter $b$, with values varying between $b\sim0.4$ and $1$, i.e., covering almost the full spectrum from purely solenoidal ($b\sim1/3$) to purely compressive driving ($b\sim1$). However, $b\ge0.4$ throughout, indicating a dominance of compressive driving modes. For most of the simulation time, we see that $b\sim0.4-0.5$, with a dominance of compressive driving modes ($b>0.4$) leading up to the peak of the star formation event around $320-330\,\Myr$. In particular, the maximum in $\sigsfr$ follows the maximum in $b\sim1$ (nearly fully compressive driving) with a time delay of $\sim10\,\Myr$, which is approximately the turbulent turnover time for WNM gas on $\sim100\,\pc$ scales. A similar effect is seen towards the end of the simulation, where a rise in $\sigsfr$ follows shortly after a strong increase in $b$.

Thus, we find evidence that high star formation rates follow shortly ($\sim10\,\Myr$) after $b$ increases to strongly compressive driving ($b>0.5$), consistent with idealised simulations \citep{FederrathKlessen2012}, while star formation and feedback subsequently drive up the turbulent velocity fluctuations ($\mach$), which leads to a reduction in $b$ down to values associated with a natural mixture of driving modes ($b\sim0.4$) around $350-360\,\Myr$, about $20\,\Myr$ after the main star formation event.

\subsection{Results in the context of previous studies on the turbulence driving mode} \label{sec:context}

\begin{figure*}
\centering
\includegraphics[width=0.8\linewidth]{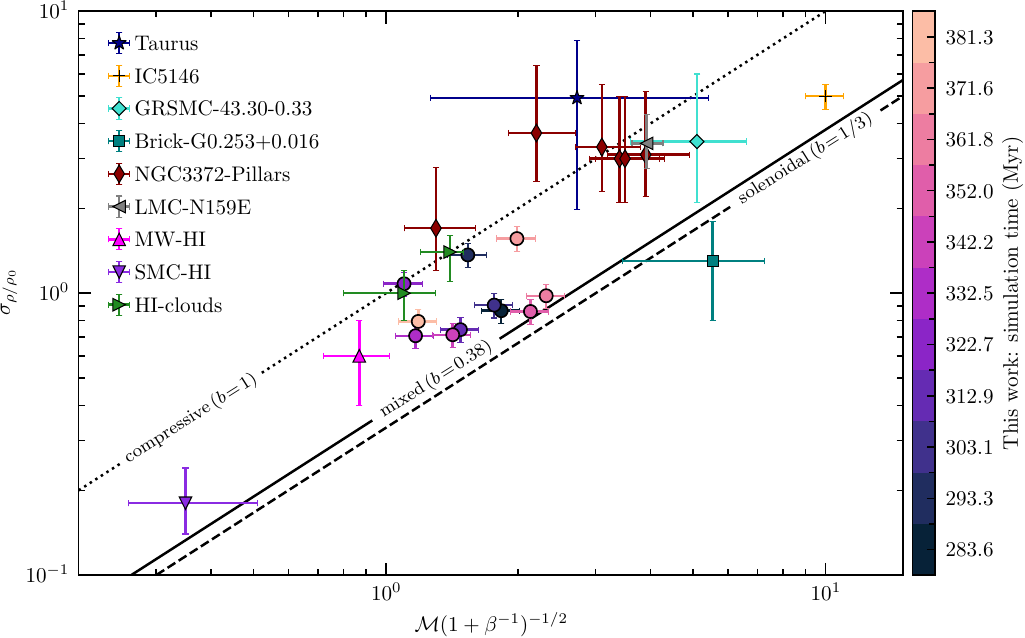}
\caption{Turbulent density dispersion -- Mach number relation for different observations, together with the present TIGRESS simulation results (shown as filled circles for different times -- see colour map). The y-axis shows the volume density dispersion ($\sigr$), and the x-axis shows the turbulent Mach number ($\mach$), including the plasma $\beta$ (ratio of thermal to magnetic pressure) factor in Eq.~(\ref{eq:b}). The three diagonal lines show the theoretical limits for compressive ($b=1$, dotted), mixed ($b=0.38$, solid), and solenoidal ($b=1/3$, dashed) driving of the turbulence \citep{FederrathDuvalKlessenSchmidtMacLow2010}. The filled circles show the 11~available snapshots of the TIGRESS simulation, coloured by the simulation time (see colour bar). The error bars on the simulation data quantify the variation in all derived turbulence parameters when comparing subsets of half the volume of the original WNM content. For context we include a variety of sources from the literature: Taurus (dark blue star) \citep{Brunt2010}, which includes magnetic field estimates and revised Mach number estimations from \citet{KainulainenTan2013}, using $^{13}$CO line imaging observations; IC5146 (orange cross) \citep{PadoanJonesNordlund1997}, using $^{12}$CO and $^{13}$CO observations; GRSMC~43.30-0.33 (turquoise diamond) \citep{GinsburgFederrathDarling2013}, observed in H$_{2}$CO absorption and $^{13}$CO emission; `The~Brick' (G0.253+0.016, teal square) \citep{FederrathEtAl2016} in the central molecular zone, using HNCO observations; `The Pillars of Creation' (NGC~3372 pillars, dark red diamonds) \citep{MenonEtAl2021}, from $^{12}$CO, $^{13}$CO and C$^{18}$O; and `The Papillon Nebula' (LMC~N159E, grey left-pointing triangle) \citep{ShardaEtAl2022}, again in $^{12}$CO, $^{13}$CO, and C$^{18}$O. Most relevant in the present context of the WNM are the $\hi$ WNM observations in the Milky Way (magenta up-pointing triangle) \citep{MarchalMivilleDeschenes2021}; the SMC (violet down-pointing triangle) \citep{GerrardEtAl2023}; and two extra-planar $\hi$ clouds (green right-pointing triangles) \citep{GerrardEtAl2024}.}
\label{fig:context}
\end{figure*}

Here we briefly compare our present results with all known observational measurements of the (magnetised) density dispersion -- Mach number relation, which is shown in Fig.~\ref{fig:context}. The most direct comparison with the present WNM simulation analysis is provided by the three observational measurements in the WNM available so far: the high-latitude WNM in the Milky Way \citep[$b\sim0.7$;][]{MarchalMivilleDeschenes2021}, the SMC \citep[$b\sim0.5$;][]{GerrardEtAl2023}, and two high-latitude clouds \citep[$b\sim1$;][]{GerrardEtAl2024}. Our measurements of $b$ in the TIGRESS simulations agree reasonably well with those observational measurements, albeit with slightly higher values of $\sigr$ and $\mach$, and noting that we include the contribution of the magnetic field ($\beta$ term in Eq.~\ref{eq:b}), while the observational studies of $\hi$ did not have access to magnetic-field data. We find that the turbulence driving parameter derived in the present study covers a large range of values, all with $b\ge0.4$, broadly consistent with the variety of observational constraints in WNM gas. This is further evidence that the dominant driving mode of WNM turbulence is on the compressive end of the spectrum ($b>0.4$), but can vary substantially in space and time.


\section{Discussion and conclusions} \label{sec:con}

In this study we presented a time-evolution analysis of the turbulence in the TIGRESS galaxy simulation framework of a solar-neighbourhood-like environment. We find time-dependent correlations between the turbulence, in particular the turbulence driving parameter $b$ (cf.~Eq.~\ref{eq:b}), and the star formation rate and associated feedback. 

On one hand, highly-compressive driving events ($b>0.5$) are followed by high star formation rates, with a delay of about $10\,\Myr$, the turbulent turnover time of WNM gas on $100\,\pc$ scales. On the other hand, the delayed feedback ($\sim20\,\Myr$ after the star formation peak) from type-B SN explosions drives up the Mach numbers in the WNM to $\mach\sim2-3$ (while $\mach\sim1-2$ during high star formation activity), which leads to increased WNM fractions and density fluctuations, lower plasma $\beta$, and a reduction of turbulence driving parameter to values of $b\sim0.4-0.5$.

Overall, for the entire time evolution of the simulation (about half a galactic orbital time), the turbulence driving mode is predominately compressive ($b>0.4$), especially during the phase leading up to the peak of star formation events when $b$ is as high as $0.8-1$. The time-averaged value of the driving parameter is $b\sim0.5$.

The methods developed here can be used in future work to investigate the turbulence in other simulation setups, including variations of the original TIGRESS model, including simulations with cosmic-ray transport, different radiation feedback types, and different metallicities. Furthermore, one can extend the work to different phases of the ISM, and produce synthetic observations for better comparisons with observational data. 


\section*{Acknowledgements}

We thank Chang-Goo Kim for helpful discussions and contributions to Sec.~\ref{sec:sims}, and for providing access to the TIGRESS data and guidance on its usage. I.A.G.~would like to thank the Australian Government and the financial support provided by the Australian Postgraduate Award. C.F.~acknowledges funding by the Australian Research Council (Discovery Projects~DP230102280 and DP250101526), and the Australia-Germany Joint Research Cooperation Scheme (UA-DAAD). C.F.~further acknowledges high-performance computing resources provided by the Leibniz Rechenzentrum and the Gauss Centre for Supercomputing (grants~pr32lo, pr48pi, and GCS Large-scale project~10391), the Australian National Computational Infrastructure (grant~ek9) and the Pawsey Supercomputing Centre (project~pawsey0810) in the framework of the National Computational Merit Allocation Scheme and the ANU Merit Allocation Scheme.


\section*{Data Availability}

The TIGRESS simulation data are available at \url{https://princetonuniversity.github.io/astro-tigress/intro.html}. The code to produce the present results is available at \url{https://bitbucket.org/chfeder/tigress_turbulence_analysis}.



\appendix

\section{Numerical resolution study} \label{app:resolution}

\begin{figure}
\centering
\includegraphics[width=0.95\linewidth]{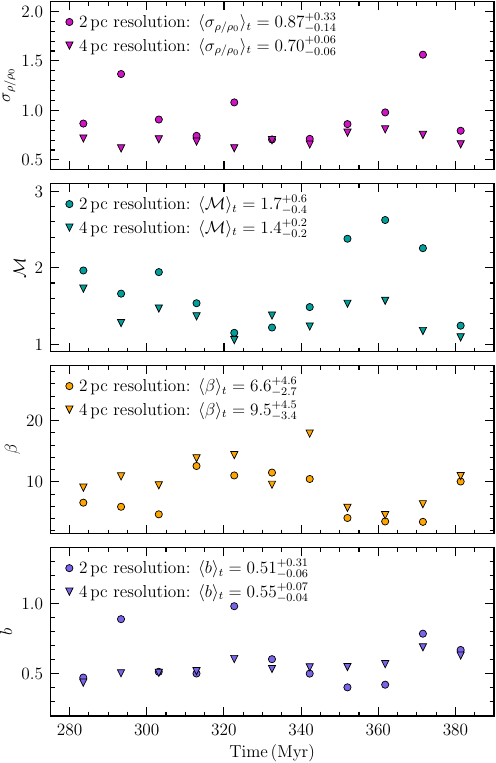}
\caption{Same as panels 3--6 in Fig.~\ref{fig:tevol}, but in addition to the $2\,\pc$-resolution case, which is the standard TIGRESS simulation case presented in the main text (shown as circles), a lower-resolution ($4\,\pc$) simulation case is shown as triangles. The case with $2\,\pc$ cell size resolves somewhat finer turbulent fluctuation, hence $\sigr$ and $\mach$ are somewhat higher and $\beta$ somewhat lower than in the case with $4\,\pc$ cell size. The resulting turbulence driving parameter $b$ (bottom panel), however, is not strongly dependent on numerical resolution. The respective time-averaged values of all quantities are shown in the legend of each panel.}
\label{fig:resolution}
\end{figure}

Figure~\ref{fig:resolution} compares the standard ($2\,\pc$ cell size) TIGRESS simulation with a lower-resolution, $4\,\pc$ run. As expected, resolving finer turbulent fluctuations leads to increases of about $20-25\%$ in $\sigr$ and  $\mach$, while the plasma $\beta$ decreases by roughly 30\%. Despite these resolution-dependent shifts in the underlying dynamical quantities, the resulting turbulence driving parameter $b$ (Eq.~\ref{eq:b}) remains relatively stable against changes in the numerical resolution, with time-averaged values of $b\approx0.51$ and $0.55$ for the $2$ and $4\,\pc$ resolutions, respectively. This demonstrates that $b$ is largely insensitive to numerical resolution within the tested (available) range for TIGRESS.

\section{Dependence on the size of the turbulence kernel} \label{app:kernelsize}

\begin{figure}
\centering
\includegraphics[width=0.95\linewidth]{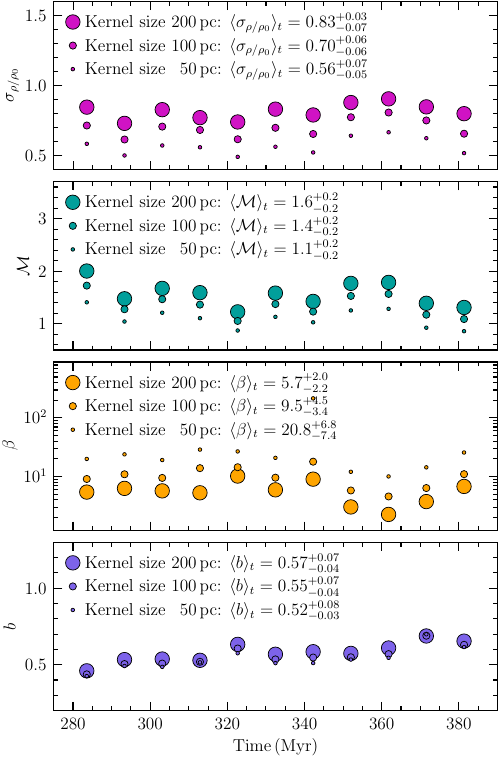}
\caption{Same as Fig.~\ref{fig:resolution}, but for different turbulence kernel sizes with FWHM of $50\,\pc$, $100\,\pc$ (default), and $200\,\pc$, at $4\,\pc$ resolution. Larger kernels probe turbulent fluctuations on larger scales, leading to an increase in $\sigr$ and $\mach$, and a decrease in $\beta$. The turbulence driving parameter $b$, however, is largely insensitive to whether a kernel size of $50$, $100$, or $200\,\pc$ is chosen.}
\label{fig:kernelsize}
\end{figure}

Figure~\ref{fig:kernelsize} shows the same as Fig.~\ref{fig:resolution} at $4\,\pc$ resolution, but for turbulence kernel sizes of $50\,\pc$, $100\,\pc$ (default), and $200\,\pc$. Increasing the kernel size probes larger-scale turbulent fluctuations, resulting in increases of about $10-20\%$ in $\sigr$ and $\mach$, accompanied by a decrease in $\beta$ by $40-100\%$ on doubling the kernel size. Despite these changes in the underlying quantities, the resulting turbulence driving parameter $b$ varies only weakly, differing by merely $\pm5\%$ for kernel sizes of $50$ and $200\,\pc$ relative to the default $100\,\pc$ case. This confirms that $b$ is fairly robust against reasonable choices of the turbulence kernel size.

\section{Magnetic-field -- density relation} \label{app:B_rho}

\begin{figure}
\centering
\includegraphics[width=0.99\linewidth]{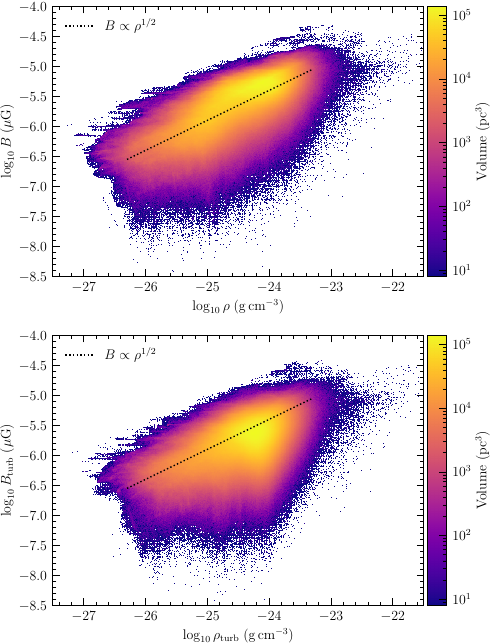}
\caption{Volume-weighted histogram of the relation between the magnetic field ($B$) and the gas density ($\rho$) at simulation time $t=293\,\Myr$, for the WNM gas before (top) and after turbulence isolation (bottom). The dotted line shows $B\propto\rho^{1/2}$, which is assumed for the plasma-$\beta$ correction term used in Eq.~(\ref{eq:b}) -- see \citet{MolinaEtAl2012}.}
\label{fig:B_rho}
\end{figure}

Figure~\ref{fig:B_rho} shows the magnetic field -- density relation at $t=293\,\Myr$, for the WNM gas before (top) and after turbulence isolation (bottom). We find that the $B\propto\rho^{1/2}$ relation (dotted line) provides a reasonable approximation to the simulation data. This is the relation used for the magnetic-field correction via plasma $\beta$ in Eq.~(\ref{eq:b}), as derived by \citet{MolinaEtAl2012}.

\section{Structure of turbulent fields} \label{app:turb_fields}

\begin{figure}
\centering
\includegraphics[width=0.95\linewidth]{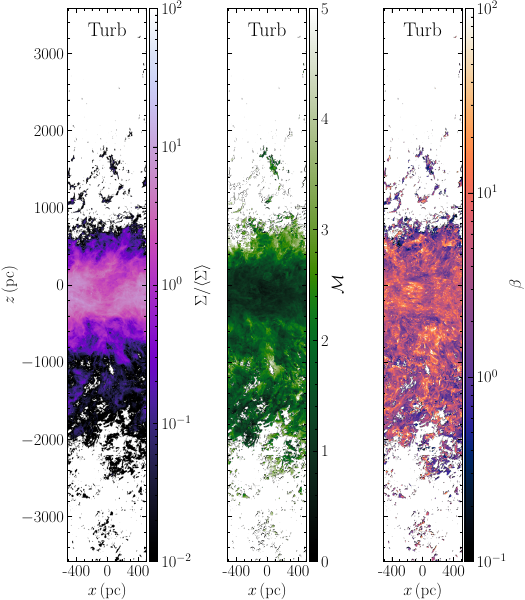}
\caption{Similar to Fig.~\ref{fig:simbox}, but for the turbulent WNM components of the density (left), Mach number (middle), and plasma $\beta$ (right), i.e., after the turbulence-isolation procedure (cf.~Sec.~\ref{sec:turbiso}).}
\label{fig:turb_fields}
\end{figure}

Figure~\ref{fig:turb_fields} shows the turbulent components of the density (left), Mach number (middle), and plasma $\beta$ (right) after the turbulence-isolation procedure (cf.~Sec.~\ref{sec:turbiso}). The 3D turbulent density field is shown as a column density, i.e., the line-of-sight (LOS) integrated turbulent density field relative to the average column density. The Mach number and plasma $\beta$ fields are shown as the mass-weighted average of the respective turbulent 3D field along the LOS. These fields represent the turbulent components on which the turbulence driving parameter estimation is carried out. We note that the 3D fields were filtered to remove large-scale fluctuations on scales $\gtrsim100\,\pc$ (cf.~Sec.~\ref{sec:turbiso}). The remaining vertical structure seen here is a result of the LOS distribution of WMN gas, i.e., there is less WNM further away from the galaxy disc midplane than in the plane, which is reflected in the projections of the turbulent components. However, the 3D fields themselves do not have this large-scale ($\sim\mathrm{kpc}$) vertical structure --- this merely arises in projection, i.e., there is more WNM gas along lines-of-sight through the disc midplane than further away from it.

\section{Instantaneous correlation of sound speed and velocity dispersion with star formation} \label{app:turbsfrcorr_cs_sigv}

\begin{figure}
\centering
\includegraphics[width=0.95\linewidth]{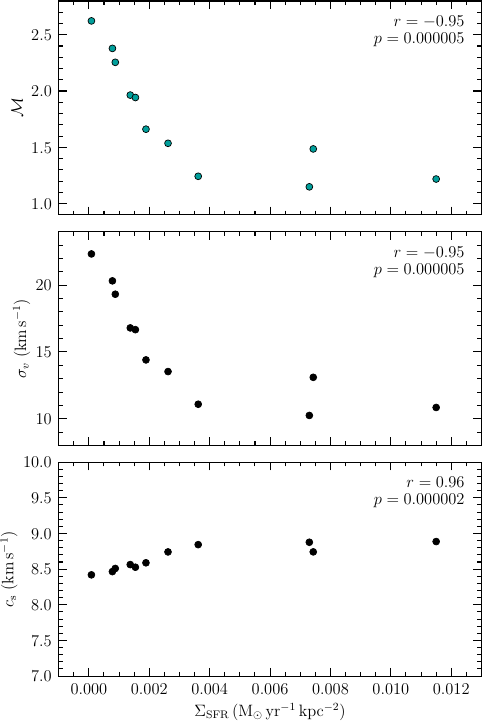}
\caption{As Fig.~\ref{fig:turbsfrcorr}, but for the turbulent Mach number ($\mach$, repeated for comparison, top panel), the turbulent velocity dispersion ($\sigv$, middle panel), and the sound speed ($\cs$, bottom panel). We see that $\mach$ is practically determined by $\sigv$, as $\cs$ stays nearly constant to within 5\%.}
\label{fig:turbsfrcorr_cs_sigv}
\end{figure}

Figure~\ref{fig:turbsfrcorr_cs_sigv} shows the same as Fig.~\ref{fig:turbsfrcorr}, but only for the turbulent Mach number (top panel), and in addition we show the turbulent velocity dispersion (middle panel), and the sound speed (bottom panel), i.e., the physical quantities that determine the Mach number, $\mach=\sigv/\cs$. This demonstrates that the Mach number is almost exclusively driven by changes in the velocity dispersion rather than the sound speed. Indeed, the sound speed is nearly constant throughout, and only increases by 5\% between low and high $\sigsfr$, as a consequence of the WNM phase selection (cf.~Sec.~\ref{sec:wnm}), which can be approximated as nearly isothermal.

\bsp	
\label{lastpage}
\end{document}